\documentclass[twocolumn,showpacs,pra]{revtex4}

\usepackage{amsmath,amssymb,graphicx,psfrag,color}

\begin{document}
\title{
Linear optics for direct observation of quantum violation of pigeonhole principle by joint weak measurement
}
\author{Kazuhiro Yokota}
\affiliation{Graduate School of Science, Osaka University, Toyonaka, Osaka 560-8531, Japan}
\author{Nobuyuki Imoto}
\affiliation{Graduate School of Science, Osaka University, Toyonaka, Osaka 560-8531, Japan}


\begin{abstract}
When three pigeons are in two pigeonholes, at least two of them should be in the same pigeonhole, which is called pigeonhole principle.
Recently Aharonov {\it el al} have shown that the principle can be violated in a pre-postselected quantum system.
This violation sheds light on a new aspect of quantum correlations, as both of the pre and postselected states are separable states.
To prove this kind of quantum correlations, a measurement apparatus must be arranged in entanglement to perform a joint measurement on the quantum system.
In this paper we discuss how to design the state of entangled meters to achieve such a measurement.
We also propose an experimental setup composed of linear optics for direct observation of violation of pigeonhole principle.
\end{abstract}

\maketitle

\section{Introduction - Quantum violation of pigeonhole principle}
\label{sec:introduction}
In quantum mechanics a counter-intuitive phenomenon often appears.
For instance, when we consider a two-qubit system, they can show a quantum correlation which cannot be achieved by a classical correlation.
One of the famous examples is violation of Bell inequality, which has clarified the concept of entanglement, namely non-separability of a quantum state \cite{Bell1, Bell2}.
However, an entangled state is not always needed to reveal such a quantum correlations, especially, in the case when a certain pair of initial and final states of a quantum system is chosen (i.e. pre-postselection) \cite{Po1}. 
Actually Aharonov {\it el al} have recently shown violation of pigeonhole principle (VPP) in which a new aspect of quantum correlations is uncovered in pre and postselected state, both of which are interestingly separable states \cite{Pigeon}; Our goal in this paper is to propose a practical experiment for observation of VPP.

Pigeonhole principle says that when three pigeons are in two pigeonholes, at least two of them are expected to be found in the same pigeonhole.
Although this statement seems to be trivial, it can be violated when a pigeon is allowed to take a pre-postselected quantum state.

In the context of optics, we would like to review the story.
Consider three photons numbered as I, I\hspace{-.1em}I, and I\hspace{-.1em}I\hspace{-.1em}I, which can go through two paths $|0\rangle$ and $|1\rangle$ respectively.
Since a photon passes $|0\rangle$ or $|1\rangle$, we will infer from pigeonhole principle that at least two of the three photons take the same path.
To be more accurate, at least one of the observables, $\hat\Pi^{\rm same}_{\rm XY}=|00\rangle_{\rm XY}\langle 00|+|11\rangle_{\rm XY}\langle 11|$, is expected to turn the value of $1$, where XY = \{I II\hspace{-.1em}, I\hspace{-.1em}I I\hspace{-.1em}I\hspace{-.1em}I, I\hspace{-.1em}I\hspace{-.1em}I I\}.

\begin{figure}
  \begin{center}
	 \includegraphics[scale=0.9]{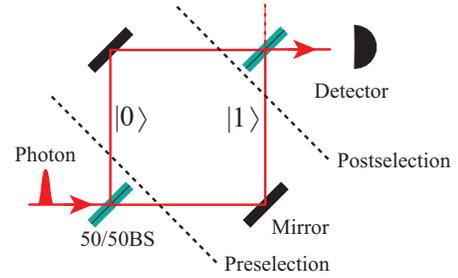}
  \end{center}
  \caption{Mach-Zehnder interferometer.
50/50BS represents a beam splitter with transmissivity/reflectivity of 50\%. 
A photon can take two paths, $|0\rangle$ and $|1\rangle$.
By adjusting the path lengths, we can perform a pre-postselection on the path state.
}
\label{fig:mz}
\end{figure}

Then we consider the case that each photon takes a superposition on the path state as shown in Fig.\ref{fig:mz}; 
We suppose that Mach-Zehnder interferometer is prepared for each photon.
The difference of the path lengths in the interferometer is adjusted so that an incident photon evolves to $(|0\rangle +|1\rangle)/\sqrt{2}$ and is finally postselected in $(|0\rangle +i|1\rangle)/\sqrt{2}$, when it appears at the port of the detector.
As a result, the initial state and the final state of three photons are given by $|i\rangle =(|0\rangle +|1\rangle)^{\bigotimes 3}/\sqrt{2^3}$ and $|f\rangle =(|0\rangle+i|1\rangle)^{\bigotimes 3}/\sqrt{2^3}$ respectively.
Note that both of them are separable states.

In this setup, we would like to ask whether a pair of photons are in the same path.
When we perform such a (strong) measurement between the pre-postselection, the probability to find a pair of photons in the same path is given by Aharonov-Bergmann-Lebowitz formula \cite{ABL} as follows,
\begin{eqnarray}
P_{\rm XY}({\rm same}) = \frac{|\langle f|\hat\Pi^{\rm same}_{\rm XY}|i\rangle |^2}{|\langle f|\hat\Pi^{\rm same}_{\rm XY}|i\rangle |^2+|\langle f|\hat\Pi^{\rm diff}_{\rm XY}|i\rangle |^2} = 0,     \label{eq:ABL}
\end{eqnarray}
for any pairs of photons XY, where $\hat\Pi^{\rm diff}_{\rm XY}=\hat{1}-\hat\Pi^{\rm same}_{\rm XY}=|01\rangle_{\rm XY}\langle 01|+|10\rangle_{\rm XY}\langle 10|$.
That is to say, any two photons (pigeons) are not in the same path (pigeonhole), which is violation of pigeonhole principle (VPP) \cite{Pigeon}.

The most important thing in this discussion is that we must not ask the probability for each photon like $|0\rangle_{\rm X}\langle 0|$; not even determine which same path a pair of photons takes (i.e. $|00\rangle_{\rm XY}\langle 00|$ or $|11\rangle_{\rm XY}\langle 11|$).

Although the result of Eq.(\ref{eq:ABL}) is always obtained for any pair of photons, they cannot be verified simultaneously for all the $\hat\Pi^{\rm same}_{\rm XY}$.
The zero probability in Eq.(\ref{eq:ABL}) comes from that the numerator is zero, in other words, the state after the measurement, $\hat\Pi^{\rm same}_{\rm XY}|i\rangle$, is orthogonal to the postselection, $|f\rangle$.
Then, from the viewpoint of experiment, Eq.(\ref{eq:ABL}) represents the zero success probability of the postselection.
If we measure, for example, the pair of ${\rm I\hspace{-.1em}I}$ and ${\rm I\hspace{-.1em}I\hspace{-.1em}I}$ after the measurement of ${\rm I}$ and ${\rm I\hspace{-.1em}I}$, the state changes to $\hat\Pi^{\rm same}_{\rm I\hspace{-.1em}I I\hspace{-.1em}I\hspace{-.1em}I} \hat\Pi^{\rm same}_{\rm I I\hspace{-.1em}I}|i\rangle$, which is not orthogonal to the postselection, $|f\rangle$.
As a result, VPP is not satisfied due to the disturbance by the two measurements.

However, by using weak measurement \cite{W1, W2}, all the measurements on $\hat\Pi^{\rm same}_{\rm XY}$ corresponding to Eq.(\ref{eq:ABL}) can be simultaneously performed.
Weak measurement achieves measurement of an observable, $\hat{A}$, without disturbance on a system, whose result is called weak value defined by
\begin{eqnarray}
\langle\hat{A}\rangle_{\bf w}=\langle f|\hat{A}|i\rangle /\langle f|i\rangle
\end{eqnarray}
for a preselection $|i\rangle$ and a postslection $|f\rangle$.
In our case the weak value of $\hat\Pi^{\rm same}_{\rm XY}$ is given as follows,
\begin{eqnarray}
\langle\hat\Pi^{\rm same}_{\rm XY}\rangle_{\bf w}=0,     \label{eq:WV_pigeon}
\end{eqnarray}
for any XY, which accords with Eq.(\ref{eq:ABL}).
Since weak measurement does not disturb the system, the weak values for all the XY can be simultaneously obtained.

When the value of strong projective measurement is 0 (1), the weak value is also 0 (1);
the converse is also correct, when the measurement results in two outcome as in Eq.(\ref{eq:ABL}), namely $\hat\Pi^{\rm same}_{\rm XY}$ or $\hat\Pi^{\rm diff}_{\rm XY}$ \cite{QP1}.
On the above discussion, the use of weak measurement is based on such an association between weak values and the results of strong measurement, even if the values seem to be mutually conflicted.
Actually weak measurement has provided an interesting approach to the issues in foundation of quantum mechanics \cite{QF1}-\cite{QF9}, especially observation of a quantum paradox so far \cite{QP1} \cite{QP2}-\cite{QP11},
while there is a case in which the verification of an anomalous feature of a pre-postselected system like the above does not always rely on weak measurement \cite{SH1, SH2}.
It also has been a beneficial measurement tool in application \cite{W3}-\cite{W10}. 
Moreover weak value itself has been found to be useful for description of a quantum phenomenon \cite{W_ph1}-\cite{W_ph7}.

Recently weak measurement has been experimentally performed in the pre-postselection addressed in VPP \cite{QF9}.
The relevant pre-postselection made the contextuality confined, which was verified as an appearance of a negative weak value.
For such a purpose, they experimentally constructed joint weak values like Eq.(\ref{eq:WV_pigeon}) from single weak values through the use of separability in both of the pre and postselected states:
The joint weak value $\langle |00\rangle_{\rm XY}\langle 00|\rangle _{\bf w}$, for example, can be calculated by the product of the single weak values, $\langle|0\rangle_{\rm X}\langle 0|\rangle _{\bf w}\langle|0\rangle_{\rm Y}\langle 0|\rangle _{\bf w}$.

However the product rule is not satisfied generally, and a joint weak value is obtained via joint weak measurement, which cannot be achieved by single weak measurements.
Note that the sum rule of joint weak values is always satisfied.
While their experiment was enough to determine all the joint weak values in VPP and to achieve their purpose, direct observation of VPP should be owing to direct observation of joint weak values.
As pointed out in the seminal paper \cite{Pigeon}, to prove the quantum correlation in VPP, a measurement apparatus should be designed so that it can detect the quantum correlation directly, which is achieved by entanglement in the measurement.
In fact joint weak measurement has been performed by entangled meters \cite{QP5}.

In this paper we discuss joint weak measurement to observe VPP directly.
Although it is clarified how to perform joint weak measurement in $n$-qubits system in the next section, all the joint weak values are not needed to confirm VPP, namely Eq.(\ref{eq:WV_pigeon}).
Aiming at only the weak values we want, we arrange joint weak measurement for observation of VPP.
This makes it possible to construct a feasible optics for the observation in section \ref{sec:Opt}.
Section \ref{sec:Con} is devoted to our conclusion.

\section{Weak measurements on qubits}
\label{sec:WM}
First, to make this paper self-contained, we review how to perform weak measurement on 1-qubit system \cite{W3}, and how an entanglement plays an important role in joint weak measurement on 2-qubits system \cite{QP5}.
After that we develop the joint weak measurement for direct observation of VPP.

\subsection{1-qubit}
\begin{figure}
  \begin{center}
	 \includegraphics[scale=0.8]{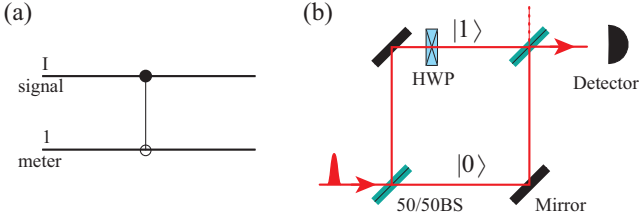}
  \end{center}
  \caption{(a)Schematic diagram for weak measurement on 1-qubit signal.
A signal and a meter qubits pass a CNOT gate.
(b)An optical setup to achieve weak measurement on 1-qubit signal.
While the path state of a photon corresponds to a signal qubit, the polarization plays a role of a meter qubit.
The half wave plate (HWP) is set so that the polarization of passing photons is changed as $H\leftrightarrow V$ to achieve a CNOT operation.  
}
\label{fig:wm}
\end{figure}

As shown in Fig.\ref{fig:wm} (a), we prepare a qubit called `meter', as a measurement apparatus for the qubit to be measured, which we call `signal.'
We assume that the signal and meter qubits are described by $|i\rangle _s =a_0|0\rangle _{\rm I}+a_1|1\rangle _{\rm I}$ and $|\xi\rangle _m=\delta|0\rangle _{1}+\varepsilon|1\rangle _{1}$ respectively, where $a_0, a_1 \in \mathbb{C}$, and $\delta, \varepsilon \in \mathbb{R}$ satisfying $0\leq\varepsilon\leq\delta\leq 1$;
The normalization is $\delta^2+\varepsilon^2=1$.
Hereafter a signal (meter) qubit is numbered with Roman (Arabic) numerals.
After passing a CNOT gate, these qubits evolve as follows,
\begin{eqnarray}
|i\rangle _s|\xi\rangle _m \
\rightarrow \ \sum_k[(\delta-\varepsilon)|k\rangle_{\rm I}\langle k|+\varepsilon]|i\rangle _s|k\rangle_{1}.   \label{eq:1wm}
\end{eqnarray} 

In the case of the optics in Fig.\ref{fig:mz}, the polarization of a photon can be used as a meter qubit.
The horizontal (vertical) polarization, $H$ ($V$), is assigned for the meter bit, $|0\rangle_1$ ($|1\rangle_1$), and the CNOT operation is achieved by placing a half wave plate (HWP) which changes the polarization as $H\leftrightarrow V$ as shown in Fig.\ref{fig:wm} (b):
The polarization (bit) is flipped, when a photon is in $|1\rangle_{\rm I}$

The correlation strength between the signal and the meter depends on $\delta$ and $\varepsilon$; the correlation strength, equivalently the measurement strength can be defined by $G=\delta^2-\varepsilon^2$.
When $G=1$ ($\delta=1, \varepsilon=0$), Eq.(\ref{eq:1wm}) results in $a_0|0\rangle _{\rm I}|0\rangle _{1}+a_1|1\rangle _{\rm I}|1\rangle _{1}$.
Such a strong correlation makes it possible to determine whether the signal is $|0\rangle_{\rm I}$ or $|1\rangle_{\rm I}$ with certainty by observing whether the meter is $|0\rangle_1$ or $|1\rangle_1$.
On the other hand, the signal and the meter are completely separable when $G=0$ ($\delta =\varepsilon$): the meter has no information about the signal, and the signal is never disturbed by the measurement on the meter, which corresponds to weak measurement.

Let us consider the case when the signal is $|0\rangle_{\rm I}$ or $|1\rangle_{\rm I}$, i.e. ($a_0=1, a_1=0$) or ($a_0=0, a_1=1$).
In $0\leq G\leq 1$, the probability when the bit of the meter, say $|k\rangle _{1}$, is the same bit of the signal, namely $|k\rangle _{\rm I}$ is given by $P_m(k)=\delta^2$, which is always larger than the probability when the bit of the meter and the signal is different (i.e. $|k\rangle _{1}$ and $|\bar{k}\rangle _{\rm I}$), $\varepsilon^2$.

We introduce the normalized readout as follows,
\begin{eqnarray}
R(k)=(P_m(k)-\varepsilon^2)/G, \label{eq:1read}
\end{eqnarray}
Then we always find $R(k)=1$ ($R(k)=0$), when the signal and the meter take the same (different) bit.
Actually when the signal takes a general state $a_0|0\rangle_{\rm I}+a_1|1\rangle_{\rm I}$, the normalized readout shows $R(k)=|a_k|^2$, i.e. $R(k)=\langle |k\rangle_{\rm I}\langle k|\rangle$.

Now let us suppose that the signal is postselected in a final state, $|f\rangle_s$.
In this case we can easily find that the normalized read out in weak measurement shows the weak value of $|k\rangle_{\rm I}\langle k|$ as follows,
\begin{eqnarray}
R(k|f)&=&(P_m(k|f)-\varepsilon^2)/G \nonumber \\
&\rightarrow& \ \langle|k\rangle_{\rm I}\langle k|\rangle_{\bf w} \ (G\rightarrow 0),
\end{eqnarray} 
where $P_m(k|f)$ represents the probability of detecting the meter as $|k\rangle_{1}$ under the condition of the successful postselection on the signal, $|f\rangle_{s}$.

\subsection{2-qubits}
\begin{figure}
  \begin{center}
	 \includegraphics[scale=0.8]{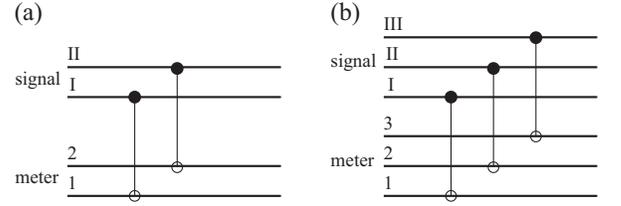}
  \end{center}
  \caption{Schematic diagram for joint weak measurement on (a)2-qubits (b)3-qubits signal.
A meter qubit is prepared for each signal qubit, and a CNOT operation is performed between these qubits.
}
\label{fig:jwm}
\end{figure}

The strategy of measurement on 1-qubit can be easily developed for 2-qubits \cite{QP5} as shown in Fig.\ref{fig:jwm} (a).
A meter qubit is prepared for each signal qubit, and they pass a CNOT gate.
The measurement strength is controlled by changing the meter state, which is given by an entangled state as follows,
\begin{eqnarray}
|\xi\rangle_m=\delta|00\rangle_{12}+\varepsilon(|01\rangle_{12}+|10\rangle_{12}+|11\rangle_{12}),   \label{eq:meter2}
\end{eqnarray}
with $0\leq\varepsilon\leq\delta\leq 1$ and the normalization, $\delta^2+3\varepsilon^2=1$.
According to the CNOT gates, the signal and the meter are correlated as follows,
\begin{eqnarray}
|i\rangle_s|\xi\rangle_m \ \rightarrow \ \sum_{kl}[(\delta-\varepsilon)|kl\rangle_{\rm I I\hspace{-.1em}I}\langle kl|+\varepsilon]|i\rangle_s|kl\rangle_{12}.
\end{eqnarray}
The measurement strength is also given by $G=\delta^2-\varepsilon^2$:
When $G=1$, the signal qubits $|kl\rangle_{\rm I I\hspace{-.1em}I}$ are strongly correlated with the meter qubits $|kl\rangle_{12}$ to be in $|kl\rangle_{\rm I I\hspace{-.1em}I}|kl\rangle_{12}$.
On the other hand, as $G\rightarrow 0$, the signal and meter gradually approximate to be separable.
Actually, with the postselection, $|f\rangle_s$, on the signal, the normalized readout, $R(kl|f)$, shows a weak value as follows,
\begin{eqnarray}
R(kl|f)&=&(P_m(kl|f)-\varepsilon^2)/G \nonumber \\
&\rightarrow& \ \langle|kl\rangle_{\rm I I\hspace{-.1em}I}\langle kl|\rangle_{\bf w} \ (G\rightarrow 0),
\end{eqnarray} 
where $P_m(kl|f)$ is the probability of detecting the meter as $|kl\rangle_{12}$ under the condition of the successful postselection on the signal.

Thus all the four joint weak values of 2-qubit, $|kl\rangle_{\rm I I\hspace{-.1em}I}\langle kl|$, can be measured, from which we can also estimate $\langle|00\rangle_{\rm I I\hspace{-.1em}I}\langle 00|\rangle_{\bf w}+\langle|11\rangle_{\rm I I\hspace{-.1em}I}\langle 11|\rangle_{\bf w}$ corresponding to $\langle\hat{\Pi}_{\rm XY}^{\rm same}\rangle_{\bf w}$ in VPP.
However, if we want to measure just $\hat{\Pi}_{\rm XY}^{\rm same}$, namely $|00\rangle_{\rm I I\hspace{-.1em}I}\langle 00|+|11\rangle_{\rm I I\hspace{-.1em}I}\langle 11|$ without determining all the joint weak values, it can be achieved by changing the state of the meter qubits but with the same setting in Fig.\ref{fig:jwm} (a);
At the cost of determination of all the joint weak values, we will be able to more accurately measure joint weak values we want, if we assume that the same amount of samples are given.
For such a purpose, the meter is given as follows,
\begin{eqnarray}
|\xi\rangle_m=[\delta(|00\rangle_{12}+|11\rangle_{12})+\varepsilon(|01\rangle_{12}+|10\rangle_{12})]/\sqrt{2},    \label{eq:meter2p}
\end{eqnarray}
where $0\leq\varepsilon\leq\delta\leq 1$ and $\delta^2+\varepsilon^2=1$.
Then the CNOT gates make the state of the signal and the meter correlated as follows,
\begin{eqnarray}
|i\rangle_s|\xi\rangle_m \rightarrow \sum_{kl}[(\delta-\varepsilon)(|kl\rangle_{\rm I I\hspace{-.1em}I}\langle kl|+|\bar k\bar l\rangle_{\rm I I\hspace{-.1em}I}\langle \bar k\bar l|) \nonumber \\
+\varepsilon]|i\rangle_s(|kl\rangle_{12}+|\bar k\bar l\rangle_{12}),
\end{eqnarray}
where $\bar 0=1$ and $\bar 1=0$.
As a result, the correlation between $|kl\rangle_{\rm I I\hspace{-.1em}I}+|\bar k \bar l\rangle_{\rm I I\hspace{-.1em}I}$ and $|kl\rangle_{12}+|\bar k \bar l\rangle_{12}$ is controlled by the measurement strength $G=\delta^2-\varepsilon^2$.
The normalized readout shows a joint weak value as follows,
\begin{eqnarray}
R(kl, \bar k \bar l|f)=(P_m(kl|f)+P_m(\bar k\bar l|f)-\varepsilon^2)/G \nonumber \\
\rightarrow \ \langle|kl\rangle_{\rm I I\hspace{-.1em}I}\langle kl|\rangle_{\bf w} +  \langle|\bar k\bar l\rangle_{\rm I I\hspace{-.1em}I}\langle \bar k\bar l|\rangle_{\bf w}\ (G\rightarrow 0),    \label{eq:WV_pigeon2}
\end{eqnarray}
with the successful postselection on the signal, $|f\rangle_s$.

In VPP, if we prepare the meter in Eq.(\ref{eq:meter2p}) for each $\hat{\Pi}_{\rm XY}^{\rm same}$, we can experimentally verify all the joint weak values in Eq.(\ref{eq:WV_pigeon}):
Eq.(\ref{eq:WV_pigeon}) is a joint weak value of 2-qubits, and can be verified with the 2-qubits meters only even in the 3-qubits signal.
However, this strategy is not efficient and redundant as we will see, since we need three 2-qubits meters.
Moreover we cannot take this strategy, if we suppose to use a polarization of a photon as a meter qubit as in Fig.\ref{fig:mz} (b).
In this case we can prepare only one meter qubit for each signal qubit.
Then we must consider joint weak measurement on 3-qubits signal with 3-qubits meter to verify all the $\hat{\Pi}_{\rm XY}^{\rm same}$ simultaneously in the next section.

\subsection{3-qubits} 
For joint weak measurement on 3-qubits signal, we also prepare 3-qubits meter and each pair of a signal qubit and a meter qubit goes through a CNOT gate as shown in Fig.\ref{fig:jwm}(b).
The state of the meter qubits resembles the one for 2-qubits in Eq.(\ref{eq:meter2}) as follows,
\begin{eqnarray}
|\xi\rangle_{m}&=&\delta|000\rangle_{123} \nonumber \\
& & \ \ +\varepsilon(1-|000\rangle_{123}\langle 000|)\sum_{klm} |klm\rangle_{123},
\end{eqnarray}
where $0\leq\varepsilon\leq\delta\leq 1$, and the normalization is $\delta^2+7\varepsilon^2=1$.
The summation of $klm$ takes all the combinations of the bits.

Generally, to measure a joint weak value of n-qubits signal, we prepare the n-qubits meter in 
\begin{eqnarray}
|\xi\rangle_{m}&=&\delta|00...0\rangle_{12..} \nonumber \\
& & \ \ +\varepsilon(1-|00...0\rangle_{12..}\langle 00...0|)\sum_{kl..}|kl..\rangle_{12..},
\end{eqnarray}
with the normalization, $\delta^2+(2^n-1)\varepsilon^2=1$.
A CNOT operation is also performed for each signal and meter qubits pair.
What we need is the probability distribution of detecting the meter qubits as $|kl...\rangle_m$, namely $P_m(kl...|f)$ under the condition of the postselection, $|f\rangle_s$, on the signal.
With the measurement strength $G=\delta^2-\varepsilon^2$, the normalized readout shows a joint weak value as follows,
\begin{eqnarray}
R(kl...|f)&=&(P_m(kl...|f)-\varepsilon^2)/G \nonumber \\
&\rightarrow& \ \langle|kl...\rangle_{\rm I I\hspace{-.1em}I I\hspace{-.1em}I\hspace{-.1em}I...}\langle kl...|\rangle_{\bf w} \ (G\rightarrow 0).
\end{eqnarray}

As mentioned in previous subsection, however, it is unnecessary to determine all joint weak values of 3-qubits signal for observation of VPP.
If the joint weak values of 3-qubits, $\langle |klm\rangle_{\rm I I\hspace{-.1em}I I\hspace{-.1em}I\hspace{-.1em}I}\langle klm|\rangle_{\bf w}+\langle |\bar k\bar l\bar m\rangle_{\rm I I\hspace{-.1em}I I\hspace{-.1em}I\hspace{-.1em}I}\langle \bar k\bar l\bar m|\rangle_{\bf w}$, are given, all the joint weak values in 2-qubits corresponding to Eq.(\ref{eq:WV_pigeon}) can be determined:
For example, $\langle\hat\Pi^{\rm same}_{\rm I I\hspace{-.1em}I}\rangle_{\bf w}=\langle |00\rangle_{\rm I I\hspace{-.1em}I}\langle 00|\rangle_{\bf w}+\langle |11\rangle_{\rm I I\hspace{-.1em}I}\langle 11|\rangle_{\bf w}=(\langle |000\rangle_{\rm I I\hspace{-.1em}I I\hspace{-.1em}I\hspace{-.1em}I}\langle 000|\rangle_{\bf w}+\langle |111\rangle_{\rm I I\hspace{-.1em}I I\hspace{-.1em}I\hspace{-.1em}I}\langle 111|\rangle_{\bf w})+(\langle |001\rangle_{\rm I I\hspace{-.1em}I I\hspace{-.1em}I\hspace{-.1em}I}\langle 001|\rangle_{\bf w}+\langle |110\rangle_{\rm I I\hspace{-.1em}I I\hspace{-.1em}I\hspace{-.1em}I}\langle 110|\rangle_{\bf w})$.
Actually, as in Eq.(\ref{eq:meter2p}), such a joint weak measurement can be achieved by using the meter qubits as follows, 
\begin{eqnarray}
|\xi\rangle_m=\Big[\delta(|000\rangle_{123}+|111\rangle_{123})+\varepsilon(1-|000\rangle_{123}\langle 000| \nonumber \\
-|111\rangle_{123}\langle 111|)\sum_{klm}|klm\rangle_{123}\Big]/\sqrt{2},   \label{eq:meter3p}
\end{eqnarray}
where $0\leq\varepsilon\leq\delta\leq 1$, and the normalization satisfies $\delta^2+3\varepsilon^2=1$.
Then the normalized readout presents the joint weak value as follows,
\begin{eqnarray}
R(klm,\bar k\bar l\bar m|f)=(P_m(klm|f)+P_m(\bar k\bar l\bar m|f)-\varepsilon^2)/G \nonumber \\
\rightarrow \ \langle|klm\rangle_{\rm I I\hspace{-.1em}I I\hspace{-.1em}I\hspace{-.1em}I}\langle klm|\rangle_{\bf w}+\langle|\bar k\bar l\bar m\rangle_{\rm I I\hspace{-.1em}I I\hspace{-.1em}I\hspace{-.1em}I}\langle \bar k\bar l\bar m|\rangle_{\bf w} \ (G\rightarrow 0).    \label{eq:3q_nr}
\end{eqnarray}

As the number of weak values to be measured is reduced, the counting data on the meter qubits can be made good use for joint weak values we want more accurately.
In addition, from the viewpoint of practical experiment, the meter qubits in Eq.(\ref{eq:meter3p}) can be prepared easily as shown in the next section.

\section{Linear optics for observing violation of quantum pigeonhole principle}
\label{sec:Opt}
The meter in Eq.(\ref{eq:meter3p}) can be rewritten as follows,
\begin{eqnarray}
|\xi\rangle_m =\Big[\sum_{xyz}[\alpha(|00\rangle_{xy}+|11\rangle_{xy})+\beta(|01\rangle_{xy}+|10\rangle_{xy})] \nonumber \\
\otimes \ (|0\rangle_z+|1\rangle_z)\Big]/\sqrt{6}, \label{eq:meter3p_2}
\end{eqnarray}
with $\delta=\sqrt{3}\alpha$ and $\varepsilon=(\alpha +2\beta)/\sqrt{3}$, where the suffixes in the summation take $xyz=\{123, 231, 312\}$.
It is easily noticed that, the term to be summed in Eq.(\ref{eq:meter3p_2}) can be decomposed to the meters we have discussed in the previous section:
While the meter qubits, $xy$, are just the same as the 2-qubits meter in Eq.(\ref{eq:meter2p}), the meter qubit, $z$, corresponds to no measurement in 1-qubit case (Eq.(\ref{eq:1wm}) with $\delta=\varepsilon$ i.e. $G=0$).
At any rate, the meter qubits in Eq.(\ref{eq:meter3p_2}) is the superposition of such 1-qubit and 2-qubits meters as $xyz=\{123, 231, 312\}$.
The measurement strength is also given by $G=\delta^2-\varepsilon^2=4(\alpha-\beta)(2\alpha+\beta)/3$, and $G\sim0$ corresponds to $\alpha\sim\beta$.

\begin{figure}[b]
  \begin{center}
	 \includegraphics[scale=1.0]{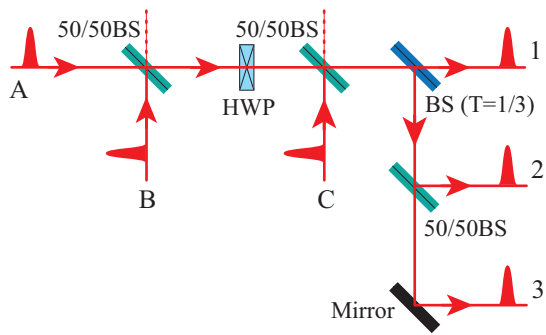}
  \end{center}
  \caption{Three photons are incident to the port A, B and C.
The polarizations of the photons going to the port A, B are entangled as ${\rm cos}\theta|HH\rangle_{AB}+{\rm sin}\theta|VV\rangle_{AB}$, while the photon for the port C is given by $(|H\rangle_C+|V\rangle_C)/\sqrt{2}$.
The angle of the optical axis of the HWP is set at 22.5 degrees:
For example, a horizontally polarized photon is changed as $|H\rangle\rightarrow (|H\rangle+|V\rangle)\sqrt{2}$ after passing the HWP.
The BS (T=1/3) represents a beam splitter with the transmission probability of $1/3$ for passing photons.
If we take only the events that one photon appears at the port 1,2 and 3 simultaneously, the polarization of the photons is given by Eq.(\ref{eq:meter_opt}).
}
\label{fig:exp}
\end{figure}

\begin{figure}[b]
  \begin{center}
	 \includegraphics[scale=0.6]{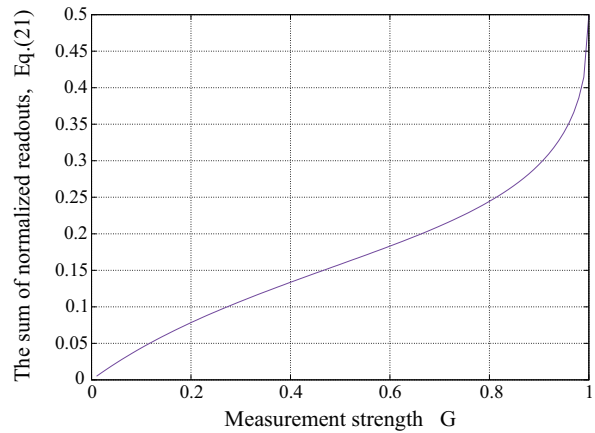}
  \end{center}
  \caption{The result of calculation of Eq.(\ref{eq:NR_pigeon}).
As the measurement strength becomes weaker, $G\rightarrow 0$, it indicates the weak value, $\langle\hat\Pi^{\rm same}_{\rm I I\hspace{-.1em}I}\rangle_{\bf w}$.
The others corresponding to $\langle\hat\Pi^{\rm same}_{\rm I\hspace{-.1em}I I\hspace{-.1em}I\hspace{-.1em}I}\rangle_{\bf w}$ and $\langle\hat\Pi^{\rm same}_{\rm I\hspace{-.1em}I\hspace{-.1em}I I}\rangle_{\bf w}$ also result in the same plots.
}
\label{fig:cal}
\end{figure}

In the case of optics, the polarization of a photon plays a role of meter qubit, and a HWP performs the CNOT operation as shown in Fig.\ref{fig:wm} (b).
Then what we need is entangled polarizations of three photons corresponding to Eq.(\ref{eq:meter3p_2}) as follows,
\begin{eqnarray}
\Big[\sum_{xyz}[\alpha(|HH\rangle_{xy}+|VV\rangle_{xy})+\beta(|HV\rangle_{xy}+|VH\rangle_{xy})] \nonumber \\
\otimes \ (|H\rangle_z+|V\rangle_z)\Big]/\sqrt{6}. \label{eq:meter_opt}
\end{eqnarray}
This entangled state can be prepared by superposing one entangled photon pair and one photon.
Actually, in Fig.\ref{fig:exp}, an entangled photon pair, ${\rm cos}\theta|HH\rangle_{AB}+{\rm sin}\theta|VV\rangle_{AB}$, is incident to the port A and B, and a photon in $(|H\rangle_C+|V\rangle_C)/\sqrt{2}$ enters the port C.
$\alpha$ and $\beta$ are related to $\theta$ as $\alpha=({\rm cos}\theta +{\rm sin}\theta)/(2\sqrt{1+2{\rm cos}^2\theta})$ and $\beta=({\rm cos}\theta -{\rm sin}\theta)/(2\sqrt{1+2{\rm cos}^2\theta})$.
By straightforward calculation, we can easily find the output photons are given by Eq.(\ref{eq:meter_opt}) as long as one photon appears at each port (1, 2, and 3), the probability of which is given by $(1+2{\rm cos}^2\theta)/72$.
Note that, after the meter preparation, each photon is incident to each Mach-Zehnder interferometer in Fig.\ref{fig:wm} (b) for observation of VPP.
Then a coincidence count of three photons assures the polarization was prepared as in Eq.(\ref{eq:meter_opt}).
Estimating the probability distributions of the polarizations of the detected three photons, we can determine the normalized readout in Eq.(\ref{eq:3q_nr}) and, finally, obtain the weak value of Eq.(\ref{eq:WV_pigeon}) for VPP.
For example, $\hat\Pi^{\rm same}_{\rm I I\hspace{-.1em}I}$ is given as follows,
\begin{widetext}
\begin{eqnarray}
& &R(HHH, VVV|f) + R(HHV,VVH)   \label{eq:NR_pigeon} \\
&=&(P_m(HHH|f)+P_m(VVV|f)+P_m(HHV|f)+P_m(VVH|f)-2\varepsilon^2)/G   \nonumber \\
&\rightarrow& \ \langle|000\rangle_{\rm I I\hspace{-.1em}I I\hspace{-.1em}I\hspace{-.1em}I}\langle 000|\rangle_{\bf w}+\langle|111\rangle_{\rm I I\hspace{-.1em}I I\hspace{-.1em}I\hspace{-.1em}I}\langle 111|\rangle_{\bf w} + \langle|001\rangle_{\rm I I\hspace{-.1em}I I\hspace{-.1em}I\hspace{-.1em}I}\langle 001|\rangle_{\bf w}+\langle|110\rangle_{\rm I I\hspace{-.1em}I I\hspace{-.1em}I\hspace{-.1em}I}\langle 110|\rangle_{\bf w}  \ \ \ (G\rightarrow 0) \nonumber \\
&\rightarrow&\langle|00\rangle_{\rm I I\hspace{-.1em}I}\langle 00|\rangle_{\bf w}+\langle|11\rangle_{\rm I I\hspace{-.1em}I}\langle 11|\rangle_{\bf w} = \langle\hat\Pi^{\rm same}_{\rm I I\hspace{-.1em}I}\rangle_{\bf w} \ \ \ (G\rightarrow 0).
\end{eqnarray}
\end{widetext}
Fig.\ref{fig:cal} shows the expected result derived from calculation: the value at $G=0$ shows  $\langle\hat\Pi^{\rm same}_{\rm I I\hspace{-.1em}I}\rangle_{\bf w}=0$.
The other joint weak values, $\langle\hat\Pi^{\rm same}_{\rm I\hspace{-.1em}I I\hspace{-.1em}I\hspace{-.1em}I}\rangle_{\bf w}$ and $\langle\hat\Pi^{\rm same}_{\rm I\hspace{-.1em}I\hspace{-.1em}I I}\rangle_{\bf w}$, are also obtained in the similar manner.

\section{Conclusion}
\label{sec:Con}
Weak value have shed light on a description of quantum nature in a different manner.
For example, weak values represent a quantum paradox very well, while the values finally circumvent the paradox.
Considering a higher dimensional system where joint weak values make the scene, we will encounter a stranger situation.
There a new aspect of quantum correlation appears, one of which is violation of pigeonhole principle, that is, a strange correlation between three separable pigeons.
For direct verification of these quantum feature, we need to build up an experimental technique of joint weak measurement.

We have discussed joint weak measurement in qubits system by using entangled meters and a local operation (CNOT gate) between a signal qubit and a meter qubit.
Moreover the meter was arranged to be specialized in observation of violation of pigeonhole principle, and a feasible optics was also proposed.
We hope our scheme of joint weak measurement contributes to the development of foundation of quantum mechanics via direct observation of joint weak values, especially to prove the distinctive feature of quantum correlations.

\begin{acknowledgments}
This work was supported by Core Research for Evolutional Science and Technology, Japan Science and Technology Agency (CREST, JST) JPMJCR1671.
\end{acknowledgments}

\end{document}